\documentclass[letterpaper,twocolumn,american,showpacs,prl,aps,superscriptaddress]{revtex4}
\usepackage[latin1]{inputenc}
\usepackage{bm}
\usepackage{multirow}
\usepackage{amssymb}
\usepackage{amsbsy}
\usepackage{amsmath}
\usepackage{stmaryrd}
\usepackage{graphicx}
\usepackage{subfigure}
\usepackage{epsfig}
\usepackage{hyperref}
\makeatletter
\usepackage{pifont}
\makeatother

\begin{document}

\title{Onset of Fokker-Planck dynamics within a Closed Finite Spin System}

\author{Hendrik Niemeyer}
\email{hniemeye@uos.de}
\affiliation{Fachbereich Physik, Universit\"at Osnabr\"uck,
             Barbarastrasse 7, D-49069 Osnabr\"uck, Germany}

\author{Daniel Schmidke}
\email{dschmid@uos.de}
\affiliation{Fachbereich Physik, Universit\"at Osnabr\"uck,
             Barbarastrasse 7, D-49069 Osnabr\"uck, Germany}

\author{Jochen Gemmer}
\email{jgemmer@uos.de}
\affiliation{Fachbereich Physik, Universit\"at Osnabr\"uck,
             Barbarastrasse 7, D-49069 Osnabr\"uck, Germany}

\date{\today}

\begin{abstract}
Relaxation according to Fokker-Planck equations is a standard scenario in classical statistical mechanics. It is however not obvious how such an equilibration may emerge within a closed, finite quantum system. We present an analytical and numerical analysis of a system comprising sixteen spins in which spatial inhomogeneities of the magnetization relax approximately in accord with a standard Fokker-Planck equation for a Brownian particle in a parabolic potential.
\end{abstract}


\maketitle
\begin{figure}[htbp]
\centering
\hspace{-1.0cm}
\includegraphics[width=9.0cm]{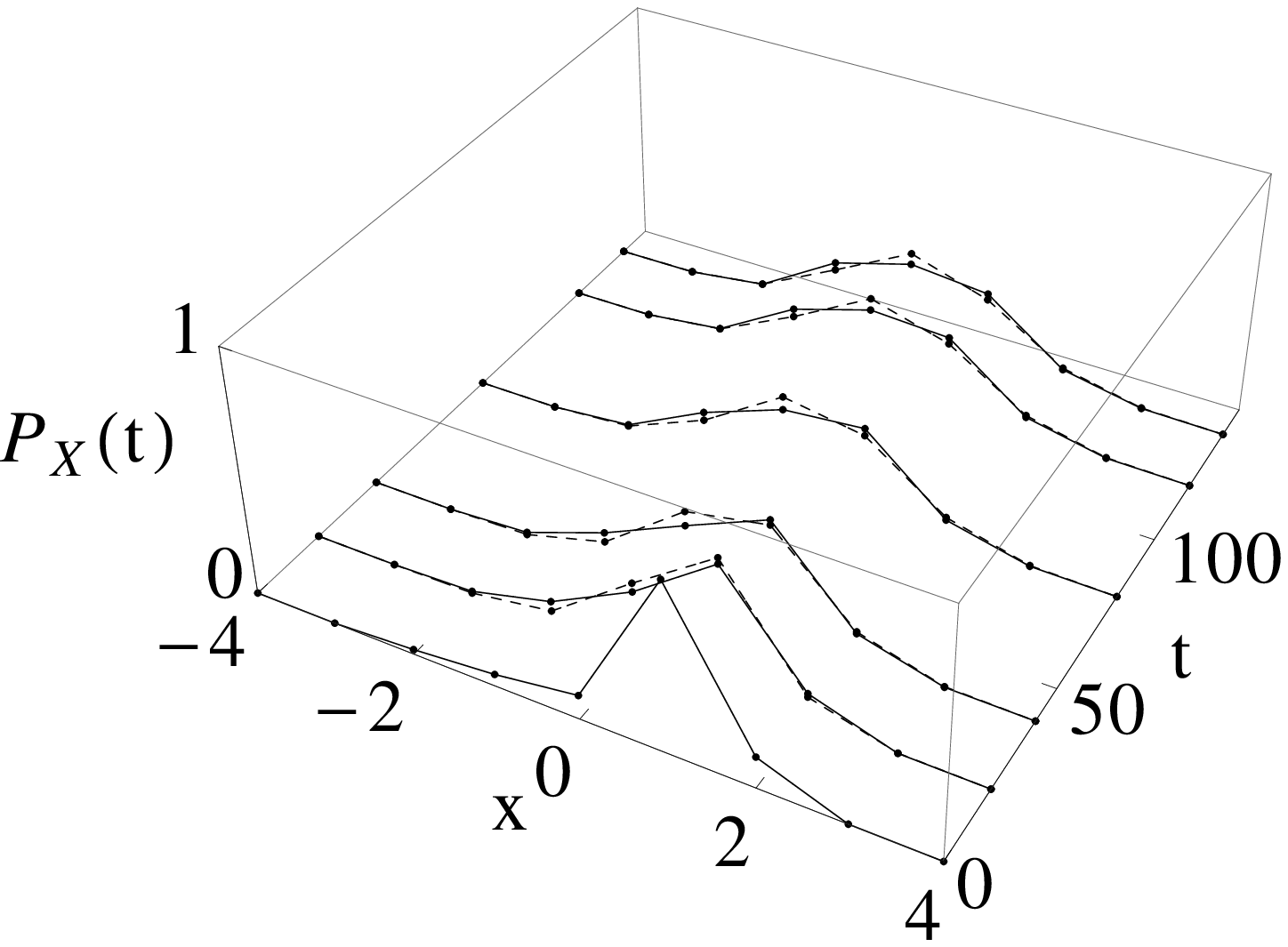}
\caption{Evolution of the probabilities of the magnetization differences $X$ for initial state $\hat{\rho}_1(0)$ (see \eqref{inist}) and coupling strength $\kappa = 0.2$. Solid lines correspond to the evolution as following from the Schroedinger equation, dashed lines to the evolution as following from the (discrete) Fokker-Planck equation (\ref{stocrat})}
\label{fancyplot}
\end{figure}
\begin{figure}[htbp]
\centering
\hspace{-1.0cm}
\includegraphics[width=9.0cm]{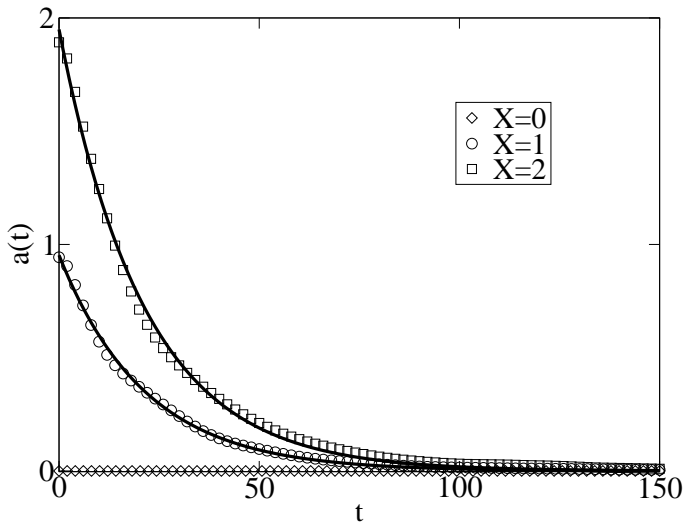}
\caption{Evolutions of the expectation values of the magnetization difference for initial states $\hat{\rho}_X(0)$ (see \eqref{inist}) and coupling strength $\kappa = 0.2$. Symbols correspond to the quantum evolutions, solid lines to the evolutions as following from the Fokker-Planck equation. } 
\label{mean}
\end{figure}
\begin{figure}[htbp]
\centering
\hspace{-1.0cm}
\includegraphics[width=9.0cm]{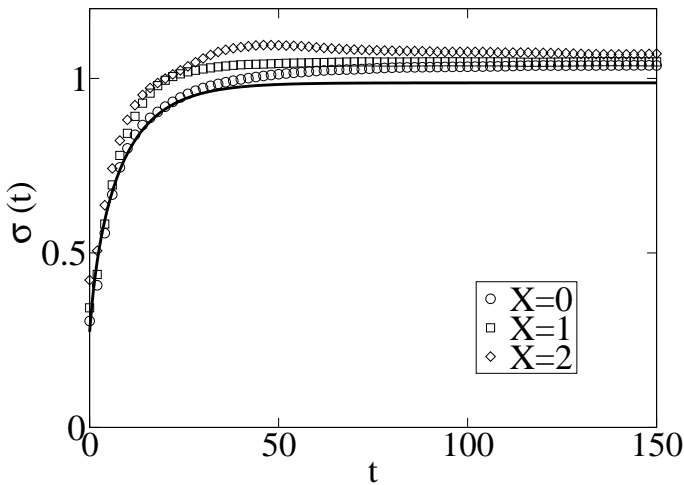}
\caption{Evolutions of the variances of the magnetization difference for initial states $\hat{\rho}_X(0)$ (see \eqref{inist}) and coupling strength $\kappa = 0.2$. Symbols correspond to the quantum evolutions, solid lines to the evolutions as following from the Fokker-Planck equation. } 
\label{varianz}
\end{figure}
\begin{figure}[htbp]
\centering
\hspace{-1.0cm}
\includegraphics[width=8.0cm]{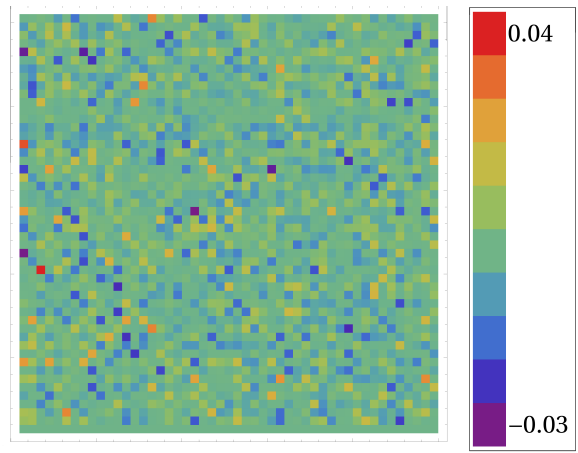}
\caption{Visualization of the fine structure of the matrix block corresponding to the $X=0 \rightarrow X=1$ transition. Displayed are the values of $50 \times 50$ matrix elements from the center of the above block w.r.t. energy. Obviously there is no apparent systematic pattern, the entries appear random} 
\label{a}

\end{figure}
\begin{figure}[htbp]
\centering
\hspace{-1.0cm}
\includegraphics[width=9.0cm]{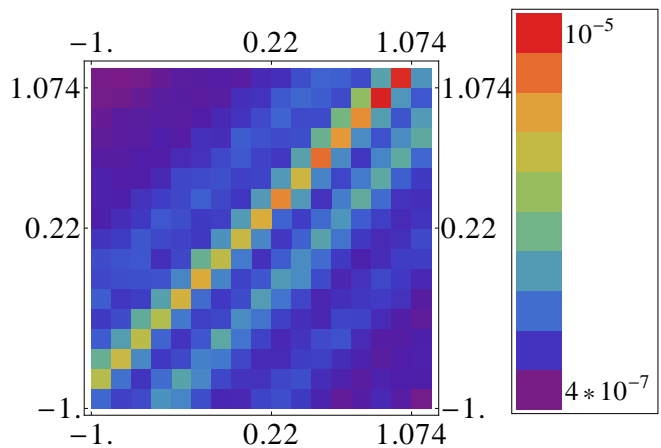}
\caption{Visualization of the coarse structure of the matrix block corresponding to the $X=0 \rightarrow X=1$ transition. Displayed are the averages over the absolute squares of the matrix elements over sectors of size $0.12 \times 0.12$ w.r.t energy. Both axis correspond to energy. Obviously there is a smooth coarse structure with a concentration of weight towards transitions between states of similar energy.}
\label{b}
\end{figure}

Do closed quantum systems show thermal equilibration and if so how does this equilibration proceed? Since the Schroedinger equation does not feature any attractive fixed point and many systems may ultimately be viewed as closed quantum systems, this question has been debated from the early days of quantum mechanics \cite{Pauli1928, Goldstein2010, Schroedinger1948, VanHove1955, Tasaki1998}. Recently this question has regained considerable attention, partially due to experimental progress in the field of ultra-cold atomic gases, which may provide the possibility of observing isolated quantum equilibration experimentally in a well controlled way \cite{Bloch2008, Trotzky2012}. On the other hand theoretical concepts have been and are being developed. Some of these approaches focus on the existence of an equilibrium for generic quantum systems and analyze properties of pure states concluding that an overwhelming majority of them features certain ``typical'' properties which are then identified as equilibrium 
properties \cite{Goldstein2010, Page1993, Goldstein2006, Popescu2006, Gemmer2003, Reimann2007}. The idea that energy eigenstates are likely to belong to this typical majority as well is now known as the eigenstate thermalization hypothesis (ETH) \cite{Goldstein2010, Deutsch1991, Srednicki1994, Rigol2008, Dubey2012}. Existence of equilibrium in specific models, such as interacting spin or interacting bosonic/fermionic model is also a subject of current research, cf., e.g.,  \cite{Eckstein2008, Genway2010, Ponomarev2011, Banuls2011, Zhang2012, Santos2012}.\\
Other approaches focus primarily on the thermalization process and aim at mapping the dynamics of crucial quantities such as certain expectation values, reduced density matrices , etc. onto (quantum) master equations, sometimes taking the form of Fokker-Planck equations (e.g., for the Caldeira Legett model, etc.). This concept has been developed in great detail for open quantum systems \cite{Breuer}, i.e., the scenario of a comparatively small system in contact with a much larger system which is usually called a bath. However, concrete examples which are not of this ``system + bath'' structure but can nevertheless be reliably shown to exhibit features of standard phenomenological equilibration seem to be rare. Or as stated in \cite{Lesanovsky2012} ``It is an intriguing question whether such a Fokker-Planck equation can actually also be found for a thermalizing and closed many-body quantum system''. This recent publication addresses a system of coupled spins in which the z-component of the magnetization is 
reported to follow a Fokker-Planck equation (FPE), close to equilibrium and with a non-linear rescaling of the magnetization. Furthermore the FPE-generator is found to  grow linear in time rather than being constant. The work at hand is aimed in a similar direction. However, we intend to present a  finite spin model in which the (unrescaled) magnetization difference between to equal parts of the system relaxes according to a FPE with an time independent generator and a diffusion term  that is almost constant w.r.t. the magnetization difference. \\
Our spin model may be described as a finite, anisotropic Heisenberg spin-ladder of length $N/2=8$. The Hamiltonian reads 
\begin{align}
\hat{H} &= \hat{H}_0 +\kappa \hat{V} \nonumber \\
 \hat{H}_0  &= \sum_{i,\alpha}^7J(\hat{\sigma}_x^{\alpha,i} \hat{\sigma}_x^{\alpha, i+1} + \hat{\sigma}_y^{\alpha,i} \hat{\sigma}_y^{\alpha, i+1} + 0.6 \; \hat{\sigma}_z^{\alpha,i} \hat{\sigma}_z^{\alpha, i+1}) \nonumber \\
  \hat{V}_0 & =\sum_{i} \kappa(\hat{\sigma}_x^{L,i} \hat{\sigma}_x^{R, i} + \hat{\sigma}_y^{L,i} \hat{\sigma}_y^{R, i} + 0.6 \; \hat{\sigma}_z^{L,i} \hat{\sigma}_z^{R, i})
\end{align}
where $J=1$ is the coupling strength along the beams (labeled by $\alpha=L,R$) and $\kappa$ the coupling strength along the rungs. Specifically we investigate $\kappa=0.2$ and  $\kappa=0.15$. (This model has been suggested in the context of relaxation dynamics in \cite{Ummethum2008}, it is non-integrable at least in the sense of not being accessible by a Bethe ansatz). The observable  of which we intend to analyze the dynamics is the magnetization difference between the beams, $\hat{x}$:
\begin{equation}
 \hat{x}=\frac{1}{2} \left(\sum_{i}\hat{\sigma}_z^{L,i}-\hat{\sigma}_z^{R,i} \right)
\end{equation}
Since the z-component of the total magnetization of the entire system $\hat{S}_z$ is conserved we may restrict our analysis to a corresponding subspace.
Here we choose $S_z$=0, which means, loosely speaking,  half of the spins point up, half down.
In the following we are going to compare the true quantum dynamics of the magnetization 
difference to the dynamics as generated by a naive, stochastic model formulated on the level of a master equation. The stochastic model is constructed as 
follows: Assume there is a rate at which mutual spin-flips, i.e., simultaneous, contrariwise flips of adjacent spins occur. Let these rates be proportional to i.) the probability with which the adjacent spins actually point in opposite directions, ii.) the square of the coupling constant between the adjacent spins. Since the rates for spin-flips within the beams are much larger than for spin-flips across the beams, one may assume that the spin-flip dynamics within the chains is at an approximate local and instantaneous equilibrium during the ongoing much slower spin-exchange dynamics between the beams. This implies that the probability for an, say, up-pointing spin on the, say right beam, is the same for every site on the beam, namely just the total number of up-pointing spins on the right beam divided by the number of beam sites. Of course the respective applies to the left beam and down-spins. Furthermore due to the above time scale separation the probabilities factorize. Following these principles the 
rates for transitions between the magnetization difference subspaces $X$ are found to be:  
\begin{equation}
\label{stocrat}
R_{(X \rightarrow X \pm 1)}=\frac{\gamma \kappa^2 N}{2}\left(\frac{1}{2} \mp \frac{2X}{N}\right)^2 
\end{equation}
The corresponding master equation  exhibits already in its finite, discrete form strong similarities with a FPE for an over-damped particle in a parabolic potential: As long as the respective probabilities $P_X$ are negligible for the extreme magnetization differences $X=\pm N/4$, the first and second moments $\overline{X}  :=\sum_XXP_X,  \overline{X^2} :=\sum_XX^2P_X$ both relax mono-exponentially with corresponding relaxation rates, $R_1=2\gamma \kappa^2$, $R_2=4(1-1/N)\gamma \kappa^2$. (The FPE would yield corresponding rates with $R_2^{FP}=2R_1^{FP}$). Regarding large $N$ and $X$ it is adequate to change to a ``magnetization difference density'' $z=X/N$ which becomes effectively continuous in the limit of large $N$. Performing a truncated Kramer-Moyal expansion w.r.t. $z$ yields
\begin{equation}
\label{fopla}
\partial_t p (t,z)= -\partial_z((-\partial_zU(z)p)+\frac{1}{2}\partial_z^2(D(z)p)+\mathcal{O}(\partial_z^3)
\end{equation}
with $U(z)=\gamma \kappa^2z^2$ and $D(z)=\gamma \kappa^2(1/4+4z^2)/N$. Up to second order in the spatial derivative (i.e., in the large length scale limit) this is very similar to the standard FPE for a Brownian particle in a quadratic potential $U(z)$. The only difference is an additional quadratic dependence of the diffusion coefficient $D(z)$ on the position. This, however, becomes irrelevant for positions sufficiently close to the equilibrium position. \\
Above the emergence of standard FPE type dynamics from the analogous stochastic model has been established. But do the quantum dynamics as generated by the Schroedinger equation (setting $\hbar =1$) indeed follow the stochastic model? This will be analyzed in the following. But before embarking upon a theoretical argument let us simply illustrate the existence of this similarity  by some numerical data.\\
The first class of initial states of which we are going to display the dynamics is constructed as follows: Let $\hat{P}_X$ be the projector comprising all states that span a subspace with a given magnetization difference $X$. Let $\hat{P}_{(E,\Delta E)}$ be a projector comprising all energy eigenstates from the interval $[E-\Delta E/2, E+\Delta E/2]$. Then we chose our initial states as
\begin{equation}
\label{inist}
\hat{\rho}_X(0)=\frac{1}{Z} \hat{P}_{(0,2)}\hat{P}_X\hat{P}_{(0,2)}, \quad Z=\text{Tr}\{\hat{P}_{(0,2)}\hat{P}_X\hat{P}_{(0,2)}\}
\end{equation}
We consider initial magnetization differences $X=0,1,2,$ and compute the corresponding evolutions of mean and variance, i.e., $a(t)=\langle \hat{x}(t) \rangle$ and $\sigma^2(t)=\langle \hat{x}^2(t) \rangle- \langle \hat{x}(t) \rangle^2$. This is simply done by numerically exact diagonalization. Fig. \ref{fancyplot} shows the evolutions of the probabilities for the various magnetization differences, i.e., $P_X(t)=\langle\hat{P}_X(t)\rangle$ for the above initial state with $X=1$. The solid lines represent the evolutions as obtained from (\ref{stocrat}). In order to analyze more initial states and see the relation to the FPE dynamics in more detail we plot more data.  In Fig. \ref{mean} the $a(t)$'s are displayed, together with the corresponding evolutions as calculated from (\ref{stocrat}) (The time-scale parameter $\gamma$ is numerically determined from a projection operator approach, see below). Obviously there is good agreement. 
 In Fig. \ref{varianz} the $\sigma(t)$'s are displayed, again together with the corresponding predictions from (\ref{stocrat}). Obviously there also is a less pronounced but still reasonable similarity.\\
Further numerical investigations (the display of which we omit here for clarity) show that the agreement of the FPE, (or (\ref{stocrat})) with quantum dynamics becomes worse if $X$ becomes larger (which is in accord with the findings in \cite{Lesanovsky2012}). The agreement also becomes worse for smaller models of the same type, i.e., $N=14, 12, ...$. This encourages the guess that the agreement may become better in the limit of $X/N \ll 1$ but $N \rightarrow \infty$
\\
Now the question arises whether the agreement  between the quantum evolution and the FPE dynamics is merely incidental.
Although we are unable to rigorously specify the class of systems and observables to which a naive FPE description of the above type applies, we provide in the following a consideration which indicates that the above specific spin system is most likely not just a singular example. This consideration is based on the time-convolutionless (TCL) projection operator method \cite{Breuer}. We may define a projection superoperator $\mathcal{P}$ by:
\begin{equation}
\label{proj}
\mathcal{P}\hat{\rho}=P_X \frac{\hat{P}_X}{d_X},\quad P_X= \text{Tr} \{\hat{P}_X \hat{\rho}\}, \quad d_X= \text{Tr} \{\hat{P}_X \}
\end{equation}
Following the TCL projection formalism yields an equation of motion for the $P_X$'s which has the form of a master equation:
\begin{equation}
\label{eqmo}
\dot{P}_{Y}=\sum_{X \neq Y} R^{TCL}_{Y, X}(t)P_X -\sum_{X \neq Y} R^{TCL}_{X, Y} P_Y
\end{equation}
(This equation technically applies only for initial states that fulfill $\mathcal{P}\hat{\rho}(0)=\hat{\rho}(0)$. However, more recent investigations imply that its applicability may be considerably wider \cite{Bartsch2009}). The rates $ R^{TCL}_{Y, X}(t)$ are given as expansions in the coupling strength $\kappa$. Readily assessable are the leading (second) order contributions which in many well-known cases suffice. They read
\begin{equation}
\label{secrat}
R^{TCL 2}_{Y, X}(t):=\int_0^t C_{Y, X}(t')dt'
\end{equation}
i.e., they are integrals over correlation functions which are given by 
\begin{equation}
\label{corr}
 C_{Y, X}(t')=\frac{\kappa^2}{d_X} \text{Tr} \{[\hat{V(t')}, \hat{P}_Y][\hat{V(0)},\hat{P}_X]\}
\end{equation}
where $\hat{V}(t')$ denotes a time dependence w.r.t. the Dirac picture, i.e., unitary dynamics as generated only be $\hat{H}_0$. A straightforward but somewhat lengthy calculation yields 
\begin{equation}
\label{tclnaive}
 C_{Y, X}(0)=\delta_{Y, X \pm 1} \frac{R_{X \rightarrow X \pm 1}}{4 \gamma}
\end{equation}
This result is rigorous and implies that the correlation functions, from which the TCL-rates are calculated  by temporal integration, feature initial values that are directly proportional to the rates as obtained from the naive stochastic description (\ref{stocrat}). This finding is furthermore independent of the details of the model, i.e., the ``left'' and  ``right'' part of the system could be any Hamiltonian structure, of course the full system must preserves the $z$-component of the magnetization. For (\ref{tclnaive}) to hold the system neither has to be a spin system, it could also be a fermionic (tight-binding) system that preserves the overall particle number. In this case the observable would be the particle number difference between a  left and right part of the system. However (\ref{tclnaive}) does not necessarily render a naive description valid. Such a validity only results if i.) the true Schroedinger evolution is well approximated by a leading order TCL approach based on the projector (\ref{proj}) and ii.) all the correlation functions $ C_{X \pm 1, X}(t')$ as given in (\ref{corr}) decay in a similar fashion and on a similar time scale independent of $X$. Otherwise the initial proportionality of the correlation functions to the naive rates  (\ref{tclnaive}) does not carry over to a proportionality of the TCL-rates to the naive rates that can be made into an equality by appropriately choosing $\gamma$. Concerning ii.) one finds numerically that for the above specific spin model the correlation functions indeed decay in  similar fashion on a similar time scale of $\tau \approx 3$ for $|X| \leq 2$. From numerical integration of the correlation functions for those small $|X|$ we find that in order to obtain $R^{TCL 2}_{X \pm 1, X}(t > \tau)\approx R_{X \pm 1, X}$ one has to choose $\gamma=0.528$.  For larger $|X|$ the dynamics of the correlation functions becomes more and more irregular (longer relaxation times, persisting oscillations) accordingly the full Schroedinger dynamics are numerically found to 
be no longer in agreement with the FPE.
Concerning  i.), it has been reported  that the  matrix that represents $\hat{V}$ w.r.t. to the basis formed by the eigenstates of $\hat{H}_0$ must feature a certain structure in order for  a leading order TCL approach to yield valid results (This being true regardless of the existence of 
so-called time scale separation, see below). The features of this structure are somewhat complex when stated formally \cite{Bartsch2008, VanHove1955}, however they are surely met for matrices the elements of which are chosen independently at random according to some distribution with mean zero. Although the Hamiltonian at hand does not contain any random numbers, $\hat{V}$ shows features of a random matrix, and thus give rise to the applicability of a leading order TCL approach. To illustrate this we display Figs. \ref{a},\ref{b}. Both are meant to give an impression of the structure of the matrix block that corresponds to the $X=0 \rightarrow X=1$ transition. Fig. \ref{a} shows the individual elements of a $50 \
\times 50$ sector of the above block taken from the center w.r.t. energy. The elements are real, zero on average and show no apparent structure, 
i.e. appear random.
To visualize the coarse structure we average the absolute squares of the matrix elements over small sectors of size $0.12 \times 0.12$ w.r.t. energy (each of these sectors still contains $\approx 1.4 \cdot 10^4$ individual elements). Obviously these weights are smoothly distributed on the coarse scale with a moderate concentration towards transitions between similar energies. In this sense the matrix is similar to a matrix the elements of which are drawn independently at random according to a distribution that only depends smoothly on the distance to the diagonal elements.

\begin{figure}[htbp]
\centering
\hspace{-1.0cm}
\includegraphics[width=9.0cm]{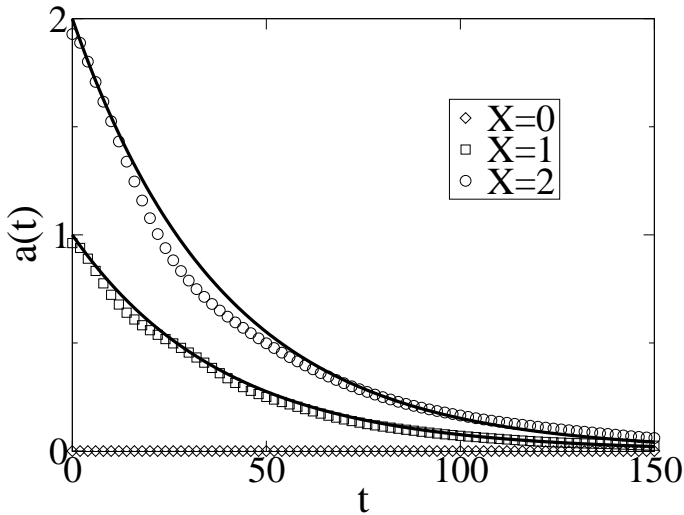}
\caption{Evolutions of the expectation values of the magnetization difference for initial states $\hat{\rho}_X(0)$ (see \eqref{inist}) and coupling strength $\kappa = 0.15$. Symbols correspond to the quantum evolutions, solid lines to the evolutions as following from the Fokker-Planck equation. } 
\label{meanw}
\end{figure}

\begin{figure}[htbp]
\centering
\hspace{-1.0cm}
\includegraphics[width=9.0cm]{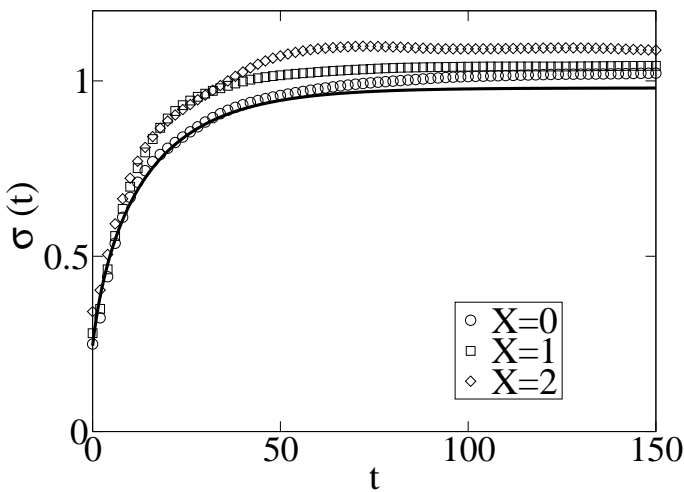}
\caption{Evolutions of the variances of the magnetization difference for initial states $\hat{\rho}_X(0)$ (see \eqref{inist}) and coupling strength $\kappa = 0.15$. Symbols correspond to the quantum evolutions, solid lines to the evolutions as following from the Fokker-Planck equation. } 
\label{varianzw}
\end{figure}

Regardless of the structure of the coupling matrix the coupling strength is of course crucial for the applicability of a leading order TCL projection approach. Generally projective approaches are considered ``weak interaction limit approaches''. And indeed in order to get essentially time-independent rates a separation of correlation and relaxation time scales is imperative. This separation grows when interactions become weaker. In our specific model for coupling strength $\kappa=0.2$ those time scales differ be a factor of $\approx 10$, i.e., time scale separation is barely implemented. This suggests even better results of a projective approach at weaker couplings. However, other  than in the context of infinite systems, for finite systems there is also a limit in the direction of weaker couplings below which results become worse \cite{quantumthermodynamics, Bartsch2008}. Numerics indicate that this limit is already reached at  $\kappa \approx 0.15$ for the model at hand as illustrated by Figs. \ref{meanw},
 \ref{varianzw}
Apparently the deviations of the true dynamics from the naive description already become more pronounced than for the case of $\kappa=0.2$. Thus for the model at hand the range of possible coupling strengths seems to be quite narrow ($\kappa$ roughly between $0.15$ and $0.2$). However, the lower limit of the coupling strengths is expected to approach zero very quickly if the system size approaches infinity. Thus the possible range of coupling strengths is expected to increase substantially for ``longer spin ladders''. Investigations in that direction are currently under way.

Concerning the applicability of the above naive stochastic description to generic systems it should be stated that it is a priori not clear whether or not the coupling $\hat{V}$ assumes the above ``quasi-random'' form when represented w.r.t. the energy eigenbasis of the decoupled parts  $\hat{H}_0$. And it is relatively simple to name examples for which this quasi-random structure does not emerge, like, e.g., any two coupled regular lattices filled with non-interacting fermions, etc. Nevertheless the quasi-random structure of $\hat{V}$ is in a mathematical sense typical, furthermore the emergence of quasi-random matrices from non-random Hamiltonians has been found in other models, e.g., \cite{Mukerjee2006, Kolovsky2004}. Thus it appears reasonable to expect that naive descriptions will apply to a much wider range of models then just the specific spin model discussed in the paper at hand. Investigations in that direction are currently under way.

An apparent feature that cannot be in accord with any TCL approach of the above type (\ref{proj}) (that features a unique fixed point) is the fact that the final variances in Figs. \ref{varianz}, \ref{varianzw} appear to depend (weakly) on the initial state. This feature is absent for the expectation values in Figs. \ref{mean}, \ref{meanw} These findings may be viewed as a direct consequence of the fact that the ETH does not perfectly apply to the observable $\hat{x}^2$: For any observable $\hat{A}$ one expects in the long run $\langle\psi|\hat{A}(t)|\psi\rangle\rightarrow \sum_{n}\langle n|\hat{A}|n\rangle|\langle n|\psi(0)\rangle|^2$ ($|n\rangle$ being energy eigenstates) if sufficiently many incommensurate energies are involved \cite{Reimann2008}. This constant value can, however, only be fully independent of the initial state if the $\langle n|\hat{A}|n \rangle$'s are independent of $n$ as the ETH claims and as is to good accuracy the case for, e.g.,  certain Hamiltonians comprising random numbers.\cite{
Cho2010, 
Dubey2012, Deutsch1991}. In the case at hand that would require $ \langle n|\hat{x}|n \rangle 
=0$ and  $ \langle n|\hat{x}^2|n\rangle=1.1$ for all eigenstates $n$ from the relevant regime. While the former holds exactly true due to symmetry, the latter is only approximately fulfilled. As a consequence the mean values truly approach the same equilibrium value regardless of the initial state while this is only approximately correct for variances. This resembles the results in \cite{Biroli2010, Gogolin2011}, which also find the possibility of some memory effects even in non-integrable systems.\\
An issue that has received considerable attention in the context of open quantum systems and irreversibility is the role of the initial state. Projection operator methods, refer routinely to specific initial states (such as factorizing, thermal, etc.) and it is not obvious within the framework of projection operator methods what kind of dynamics will result from different initial states \cite{Breuer}. The initial states we have analyzed so far are factorizing and high entropy. Now we turn towards pure, i.e., zero entropy initial states. We investigate two types, factorizing and correlated, the latter here implies entangled. To evaluate the deviation from the expected relaxation behavior we calculate $\delta= \int_0^{150}| a_Q(t)-a_{FPE}(t)|dt/150$. Here $a_Q(t)$ is the mean magnetization difference as resulting from the respective initial quantum state, $a_{FPE}(t)$ is the value as calculated from (\ref{stocrat}). For the factorizing initial states we draw pure states at random (uniformly w.r.t. the unitary 
invariant measure) but from the ``$5$ spins-up'' subspace on the left and the ``$3$ -up'' subspace on the right beam and compute the corresponding product states. For the entangled initial states we draw states from the full $X=1$ subspace at random (which are certainly entangled \cite{Page1993}) For the product and entangled states we draw five initial states for each class and average the  $\delta$'s. We find: $\delta=0.0154$ for the mixed initial state as given in (\ref{inist}) with $X=1$, $\overline{\delta}=0.0161$ for the pure product states and $\overline{\delta}=0.0162$ for the pure entangled states. Obviously all those dynamics follow the
 FPE evolution quite closely.
This supports the concept of an equilibration that proceeds almost independently of the details of the initial quantum  state as presented in \cite{Bartsch2009}.\\
We investigated the dynamics of the magnetization difference between the two beams of a specific 16 spin, ladder-type Heisenberg model. This dynamics is found to be to some extent in accord with a Fokker-Planck equation for an over-damped particle in a quadratic potential, thus making the model an example for the emergence of standard equilibration within finite closed quantum systems. An analysis which indicates that this behavior may be generic for a wider range of quantum systems is presented. 
Furthermore the equilibration appears to be in reasonable accord with the eigenstate thermalization hypothesis.  Standard equilibration is found for mixed but also pure, (to some extent random) factorizing and entangled initial states which agrees with the concept of dynamical typicality.\\
We thank R. Tumulka, S. Goldstein, J. Lebowitz and J. Ummethum for fruitful discussions on this subject.

\bibliography{spinlit}
\bibliographystyle{apsrev4-1}

\

\end{document}